\documentclass[10pt, conference]{IEEEtran}
\IEEEoverridecommandlockouts
\usepackage{cite}
\usepackage{amsmath,amssymb,amsfonts}
\usepackage{algorithmic}
\usepackage{graphicx}
\usepackage{textcomp}
\usepackage{xcolor}
\def\BibTeX{{\rm B\kern-.05em{\sc i\kern-.025em b}\kern-.08em
    T\kern-.1667em\lower.7ex\hbox{E}\kern-.125emX}}
\usepackage{booktabs}
\usepackage{multirow}

\begin{document}

\title{A Method for Network Intrusion Detection Using Flow Sequence and BERT Framework\\
\thanks{
An earlier and short version of this paper was presented at
the 18th International Conference on emerging Networking EXperiments and Technologies
Student Workshop (CoNEXT-SW '22)~\cite{own}.

This work was partly supported by JSPS KAKENHI Grant Number JP20H04172.
}
}

\author{\IEEEauthorblockN{Loc Gia Nguyen}
\IEEEauthorblockA{\textit{Graduate School of Engineering}\\
\textit{Nagaoka University of Technology}\\
Nagaoka, Niigata, Japan \\
s203145@stn.nagaokaut.ac.jp}
\and
\IEEEauthorblockN{Kohei Watabe}
\IEEEauthorblockA{\textit{Graduate School of Engineering}\\
\textit{Nagaoka University of Technology}\\
Nagaoka, Niigata, Japan \\
k\_watabe@vos.nagaokaut.ac.jp}
}

\maketitle

\begin{abstract}
A Network Intrusion Detection System (NIDS) is a tool that identifies potential threats to a network. Recently, different flow-based NIDS designs utilizing Machine Learning (ML) algorithms have been proposed as solutions to detect intrusions efficiently. However, conventional ML-based classifiers have not seen widespread adoption in the real world due to their poor domain adaptation capability. In this research, our goal is to explore the possibility of using sequences of flows to improve the domain adaptation capability of network intrusion detection systems. Our proposal employs natural language processing techniques and Bidirectional Encoder Representations from Transformers framework, which is an effective technique for modeling data with respect to its context. Early empirical results show that our approach has improved domain adaptation capability compared to previous approaches. The proposed approach provides a new research method for building a robust intrusion detection system.
\end{abstract}


\section{Introduction}
%{\bf WHY IDS ARE IMPORTANT}\\
As more of our daily lives depend on computer networks, safeguarding the information contained within these systems and maintaining service availability has become a necessity. Systems that handle financial transactions, personal information, or those that serve as social platforms are prime targets for bad actors seeking to gain financial assets or influence over society. Currently, many different cybersecurity measures are deployed in tandem to counter these threats, and intrusion detection is an important component of these measures.

%{\bf HOW DO IDS WORK}\\
Intrusion detection is the process of monitoring the events occurring in a computer system or network and analyzing them for signs of malicious activities. Network traffic is captured in either packet-based or flow-based format. Packet-based data contain complete payload information. Flow-based data are aggregated packet-based data and typically contain various properties of the collection of packets and metadata from network connections~\cite{Ring2019ASO}. NIDSs, therefore, can be separated into two categories: packet-based and flow-based. The former analyzed individual packet payloads and header information, while the latter only analyzed the aggregated properties of a collection of packets. The speed at which flow-based NIDSs analyze network traffic has made it the preferred solution for modern network security systems.

%{\bf CURRENT CHALLENGES FACED BY IDS}\\
Changes in the users' behavior or appearances of new attacks manifest as shifts in the distribution of flow features. Therefore, NIDSs need to adapt to the new benign flows while still being able to detect malicious ones. Flow-based NIDSs analyze the properties of the packet collections using machine learning algorithms, such as K-Nearest Neighbors and Random Forest. However, these machine learning algorithms often adapt poorly to changes in data distributions, which are common in real-world deployment, which in turn limits their usefulness~\cite{dart, iot}. To improve the domain adaptability of NIDSs, Camila {\it et. al.}~\cite{efc} proposed the Energy-based Flow classifier (EFC) - a statistical approach where the NIDS would model the statistical distribution of benign flows, then classifies flow that deviates from this distribution as malicious. Although it improves the domain adaptability of NIDSs, there are still limitations on certain data distributions. We assume that the reason for the limitations of current algorithms is the use of singular flows as input data, as the classifier can only model the distribution of features within a flow.

%{\bf GOAL OF THIS PAPER}\\
In this paper, we propose the use of a sequence of flows, {\it i.e.}, a collection of flows that are temporally close together, and Bidirectional Encoder Representations from Transformers (BERT) framework to obtain the representation of a network flow in a sequence. A Multi-layer Perceptron (MLP) then processes this vector representation to produce the classification result of the flow. To the best of our knowledge, this is the first example of using BERT to improve the domain adaptation capability of NIDS. This paper evaluates the performance of our proposal on the entirety of the CIDDS data sets, as opposed to a subset of data in an earlier version~\cite{own}.
%   Please explain the difference with Ref[1] below: Ref[1]: Nguyen L G,
%     Watabe K. Flow-based network intrusion detection based on BERT masked
%     language model[C]//Proceedings of the 3rd International CoNEXT Student 
%     Workshop. 2022: 7-8.

In summary, the contribution of this study is as follows.
\begin{itemize}
  \item We investigate the possibility of using information
    from a sequence of network flows to improve the domain adaptation
    capability of the network intrusion detection system.
  \item We propose an NIDS that uses BERT
  for feature extraction and MLP for classification.
  \item We demonstrate the domain adaptation capability
  of the proposed system using authentic and synthetic data sets.
\end{itemize}

The rest of the paper is organized as follows: In section II, we present an overview of prominent studies in the literature on flow-based NIDSs. Section III provides the basic concept of network flow and network intrusion detection system, the data sets used in this study, and an introduction to the BERT framework. Section IV illustrates the details of the proposal and its evaluation method. Performance evaluation results of our proposal and related discussion are provided in Section V. Finally, Section VI concludes this paper.

\section{Related Works}
The use of machine learning for implementing NIDSs has been extensively researched~\cite{survey}, with deep learning methods being employed to further improve upon machine learning methods by using deep learning models to extract features from flow data. A popular technique for feature extraction in NIDS is the autoencoder. It works on the idea of matching the output as close the to input as possible by learning the best features. Shone {\it et al.}~\cite{rf}, Yan {\it et al.}~\cite{yan} and Al-Qatf {\it et al.}~\cite{qatf} showed that autoencoders can retain important information about a flow while reducing its dimensions. This approach reduces the processing time needed by machine learning classifiers and boosts their classification capability. Andresini {\it et al.}~\cite{andresini} trained different autoencoders for each type of flow and uses the representation produced by these autoencoders to train a classifier. Khan {\it et al.}~\cite{khan} showed that autoencoders can also be used as a scoring function that takes the features of a flow as input and returns a score. The autoencoder in this case is optimized to assign a low score for normal flow and a high score for suspicious flows.

Thapa {\it et al.}~\cite{cart} conducted an experimental comparison of NIDSs based on various machine learning and deep learning techniques. Their performance was evaluated on CIDDS-001 and CIDDS-002 data sets. The study includes k-nearest neighbors, XGBoost, and decision tree classifiers as machine learning techniques. For deep learning techniques, Embedding Layer, Convolutional Neural Network, and Long Short Term Memory were considered.% The study concluded that machine learning and deep learning techniques can produce results with an accuracy of 99\% on the CIDDS dataset with a high detection rate, low false alarm rate, and relatively low training costs.

%A model is only as good as its assumptions, and in the case of NIDS models, its training data.
A model is only as good as the data used to train it. The preparation of data for training NIDSs is also an important aspect of NIDS research. Verma {\it et al.}~\cite{knn} performed a statistical analysis of the labeled flow-based CIDDS-001 dataset using k-nearest neighbor classification and k-means clustering algorithms. Abdulhammed {\it et al.}\cite{imbal} proposed several techniques to handle imbalanced training data and improve NIDSs performance. The effectiveness of sampling methods on CIDDS-001 is studied and experimentally evaluated through deep neural networks, random forest, voting, variational autoencoder, and stacking machine learning classifiers.

All of the previously mentioned papers focused on improving the detection of malicious activities and did not evaluate the performance of the approaches concerning changing data distributions. Camila {\it et al.}~\cite{efc} stated that machine learning approaches adapt poorly to changing data distributions and proposed Energy-base Flow Classifier (EFC). This algorithm is based on inverse statistics to infer a statistical model based on benign flows. The authors showed that the algorithm could adapt to different data distributions, making it suitable for detecting zero-day threats.

\section{Materials}
In this section, first, we introduce the concept of the subject of our research, network traffic flow data. Second, we present the data sets used to evaluate NIDS performance. Finally, we provide an overview of the BERT framework.

\subsection{Network traffic flow data}
A network flow is a sequence of packets carrying information between two hosts where packets have common properties. All packets within a flow share the same 5-tuple (Src IP, Src Pt, Dst IP, Dst Pt, Proto). All inter-packet times are less than an arbitrary flow expiry timeout value. Other than the 5-tuple or 3-tuple, a flow can include other flow keys, such as the number of transmitted packets, bytes, etc. There are many standards for flow data formats, including NetFlow and OpenFlow.
\subsection{Benchmark data sets}
\begin{table}
  \caption{Attributes within the CIDDS-001 and CIDDS-002 data sets}
  \label{tab:cidds}
  \resizebox{\linewidth}{!}{
  \begin{tabular}{lll}
    \toprule
\#&Name&Description\\
    \hline
1&Date first seen& Start time flow first seen\\
2&Duration&Duration of the flow\\
3&Proto& Transport Protocol (e.g. ICMP, TCP, or UDP)\\
4&Src IP&Source IP Address\\
5&Src Pt&Source Port\\
6&Dst IP&Destination IP Address\\
7&Dst Pt&Destination Port\\
8&Packets& Number of transmitted packets\\
9&Bytes& Number of transmitted bytes\\
10&Flags& OR concatenation of all TCP Flags\\
    \bottomrule
  \end{tabular}
  }
\end{table}
CIDDS-001~\cite{cidds1} and CIDDS-002~\cite{cidds2} are relatively recent intrusion detection benchmark data sets containing unidirectional NetFlow data. The CIDDS-001 data set contains two sets of flows, one is captured within a simulated OpenStack environment, and the other is captured from a server deployed on the Internet. The CIDDS-002 data set contains one set of flows from a simulated OpenStack environment.

In this study, CIDDS-001 OpenStack, CIDDS-001 External Server, and CIDDS-002 OpenStack environments are used to evaluate the domain adaptation capability of our proposal. Within the CIDDS-001 data set, a change from the simulated OpenStack environment to the external server environment is considered a domain change, as the distribution of flow features differs between the environments. Similarly, a change from CIDDS-001 to CIDDS-002 OpenStack environment is also considered a domain change. We utilize these domain changes to evaluate the domain adaptation capability of our proposal.

The network flows within CIDDS-001 and CIDDS-002 are all labeled. In the OpenStack environment of CIDDS-001 and CIDDS-002, benign flows are labeled normal, while malicious flows are labeled dos, scan, portScan, pingScan, or bruteForce, according to the type of attack. The External Server environment contains both simulated flows from the OpenStack environment and real flows from the internet. The real flows were labeled {\it unknown} for traffic to port 80 and 443 and {\it suspicious} for the remaining traffic. Flows labeled {\it unknown} are likely to be traffic from normal users and are considered benign in this study, and those labeled {\it suspicious} are considered malicious.

Table \ref{tab:cidds} shows an overview of the attributes within the CIDDS-001 and CIDDS-002 data sets. Source IP address, destination IP address, and Date first seen were not used for classification in this study as they are arbitrary and carry no useful information.

\subsection{Bidirectional Encoder Representations from Transformers}
Bidirectional Encoder Representations from Transformers (BERT)~\cite{bert} is a transformer-based machine-learning technique for NLP developed by Google. The BERT framework comprises two steps: pre-training and fine-tuning.  In pre-training, the BERT model trains on unlabeled data. For fine-tuning, the model is initialized using the pre-trained parameters and then trained using labeled data from the downstream tasks. BERT is pre-trained with two unsupervised tasks: Masked Language Modeling (MLM) and Next Sentence Prediction (NSP).  In MLM, some words in a sentence are replaced with a different token. The objective is to predict the original value of the masked words based on other unmasked words in the sentence. In NSP, BERT takes sentence pairs as input. The objective is to predict whether the second sentence in the pair is the next in the document. For fine-tuning, task-specific inputs and outputs are added to a pre-trained BERT model.

\section{Method}
In this section, we first highlight the benefit of using a sequence of flows as input for our NIDS. Next, we describe the detail of our proposed system architecture. Finally, we elaborate on the evaluation methods used in this study.

\subsection{Conditional information from flow sequence}
\begin{table}
  \caption{Example of two sequences of flows from CIDDS-001 data set}
  \label{tab:seq}
  \resizebox{\linewidth}{!}{
  \begin{tabular}{llllllll}
    \toprule
Duration&Proto&Src Pt&Dst Pt&Packets&Bytes&Flags&Class\\
\hline
\textbf{9.588}&\textbf{TCP}&\textbf{22}&\textbf{47695}&\textbf{19}&\textbf{3185}&\textbf{.AP.SF}&\textbf{suspicious}\\
\textbf{9.588}&\textbf{TCP}&\textbf{47695}&\textbf{22}&\textbf{15}&\textbf{2163}&\textbf{.AP.SF}&\textbf{suspicious}\\
\textbf{9.555}&\textbf{TCP}&\textbf{54731}&\textbf{22}&\textbf{15}&\textbf{2163}&\textbf{.AP.SF}&\textbf{suspicious}\\
\textbf{9.555}&\textbf{TCP}&\textbf{22}&\textbf{54731}&\textbf{19}&\textbf{3185}&\textbf{.AP.SF}&\textbf{suspicious}\\
\textbf{10.412}&\textbf{TCP}&\textbf{22}&\textbf{57489}&\textbf{19}&\textbf{3185}&\textbf{.AP.SF}&\textbf{suspicious}\\
\textbf{10.412}&\textbf{TCP}&\textbf{57489}&\textbf{22}&\textbf{15}&\textbf{2163}&\textbf{.AP.SF}&\textbf{suspicious}\\
0&UDP&56475&19&1&46&......&suspicious\\
0&ICMP &0&3.3&1&57&......&suspicious\\
\textbf{11.03}&\textbf{TCP}&\textbf{33056}&\textbf{22}&\textbf{19}&\textbf{2383}&\textbf{.AP.SF}&\textbf{suspicious}\\
\textbf{11.03}&\textbf{TCP}&\textbf{22}&\textbf{33056}&\textbf{21}&\textbf{3369}&\textbf{.AP.SF}&\textbf{suspicious}\\
\hline
0.054&TCP&8000&52252&7&556&.AP.SF&normal\\
0.054&TCP&52252&8000&6&515&.AP.SF&normal\\
0.077&TCP&8000&52253&7&702&.AP.SF&normal\\
0.077&TCP&52253&8000&6&586&.AP.SF&normal\\
\textbf{526.089}&\textbf{TCP}&\textbf{22}&\textbf{59862}&\textbf{178}&\textbf{24253}&\textbf{.AP.SF}&\textbf{normal}\\
\textbf{526.089}&\textbf{TCP}&\textbf{59862}&\textbf{22}&\textbf{180}&\textbf{13471}&\textbf{.AP.SF}&\textbf{normal}\\
0&TCP&53213&23&1&46&....S.&suspicious\\
0&TCP&23&53213&1&40&.A.R..&suspicious\\
0.07&TCP&8000&52243&7&702&.AP.SF&normal\\
0.07&TCP&52243&8000&6&586&.AP.SF&normal\\
    \bottomrule
  \end{tabular}
  }
\end{table}

A sequence of flows are flows that appear temporally near to one another. We can determine the appearance probability of the next flow in the network based on past flows. Table \ref{tab:seq} shows two flow sequences where a flow can be distinguished between malicious or benign based on the information provided by the sequence. The flows in bold have similar port numbers and flags, however, they are malicious in the first sequence and benign in the second sequence. If we were to classify these flow base on their features alone, these flows would all be classified as either benign or malicious. In the first sequence, these flows appear multiple times in a row and exhibit rapid changing of port numbers, this can be a pattern of an attacker retrying attacks at port 22. In the second sequence, the flows appear isolated. In cases where flow features are nearly identical, the extra information provided by examining the flows surrounding the target flow helps to distinguish them. The patterns that arise in a sequence of flows are feature agnostic, that is the same pattern can appear with arbitrary features. Therefore, information from flow sequence is robust between different domains, where features within flows might vary. In summary, with the use of sequences of flows, extra information can be made available to the classifier, and this information is independent of the features within flows.

\subsection{System architecture}
\begin{figure}
  \centerline{\includegraphics[width=\linewidth]{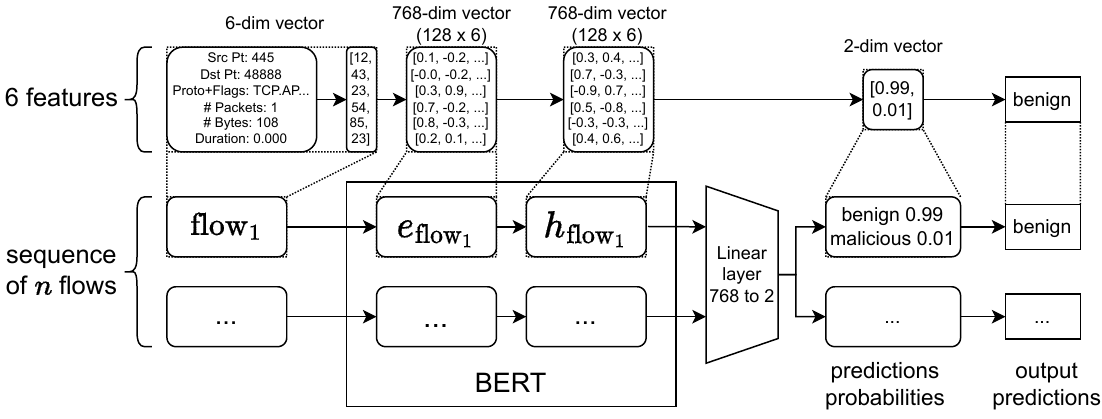}}
  \caption{Proposed system architecture}
  \label{fig:prop}
\end{figure}

We propose a method for detecting network intrusion based on the BERT and MLP models. The BERT model functions as a feature extractor, transforming network traffic flows into vectors. While the MLP model function as a classifier, classifying network traffic flows into benign and malicious flows.

%   The authors should detailedly introduce how to organize network traffic
%     flows into structures similar to natural languages.

We first organize network traffic flows into sequences, with a flow being considered analogous to a word and a sequence of flows to a sentence. Similar to how natural languages are processed, the sequences of flows are then encoded into sequences of tokens. Training of the proposed model consists of two steps. First, we pre-train the BERT model using the MLM task and only the benign flows. Since there is no inherent order to flows, we opt not to use positional encoding and special starting and ending tokens as used in the typical BERT framework. Second, we replace the MLM output layer with an MLP classifier and perform fine-tuning. In this step, we used all the flows in the data set and their labels to train the BERT model and MLP classifier. Preserving the distribution of flows within a sequence is important; therefore, the flows in the training data set are not shuffled. We use cross-entropy as the training criterion for both steps. It maximizes the likelihood of guessing the correct flow features given the context in the first step and the correct label in the second step.

\section{Evaluation}

\subsection{Model configuration}
The overall architecture of the system is illustrated in Figure~\ref{fig:prop}. Each flow contains six features, these features are encoded as numbers~($\mathrm{flow}_i$). BERT decodes each number into a 128-dimension vector, concatenates them to form a 768-dimension vector~($e_{\mathrm{flow}_i}$), and processes it to represent the flow as a different 768-dimension vector~($h_{\mathrm{flow}_i}$). The output of BERT is then passed through a Multilayer Perceptron classifier (a linear layer with softmax output), which reduces the dimension from 768 to 2. This 2-dimension vector represents the probability of each class (benign and malicious). The final prediction label is chosen to be the class with the higher probability.

\subsection{Benchmarks}
\begin{table}
  \caption{Labels within CIDDS environments}
  \label{tab:lab}
  \resizebox{\linewidth}{!}{
  \begin{tabular}{lrclrclr}
    \toprule
    \multicolumn{2}{c}{CIDDS-001 OpenStack}&&
    \multicolumn{2}{c}{CIDDS-001 External Server}&&
    \multicolumn{2}{c}{CIDDS-002 OpenStack}\\
    Label&\#&&
    Label&\#&&
    Label&\#\\
    \cline{1-2}
    \cline{4-5}
    \cline{7-8}
    {\it normal}&28051906& &{\it normal}&134240&     &{\it normal}&15598543\\
    {\it dos}&2959027&     &{\it unknown}&77923&     &{\it scan}&562640\\
    {\it portScan}&265918& &{\it suspicious}&437911&\\
    {\it pingScan}&6090&   &{\it portScan}&18719&&\\
    {\it bruteForce}&4992& &{\it bruteForce}&2448&&\\
    \bottomrule
  \end{tabular}
  }
\end{table}
\begin{table}
  \caption{Number of labels in each data set used in this study}
  \label{tab:dat}
  \resizebox{\linewidth}{!}{
  \begin{tabular}{lrrrr}
    \toprule
    &CIDDS-001
    &CIDDS-001
    &CIDDS-001
    &\multirow{2}*{CIDDS-002}\\
    &internal
    &balanced
    &external
    &\\
    \hline
    Benign&28051906
    &3236027
    &212163
    &15598543\\
    Malicious&3236027
    &3236027
    &459078
    &562640\\
    \hline
    Total&31287933
    &6472054
    &671241
    &16161183\\
    \bottomrule
  \end{tabular}
  }
\end{table}

We compare our proposal with EFC - an approach that aims at better domain adaptability. Other ML-based NIDS methods included in this comparison are Decision Tree (DT), K-Nearest Neighbors (KNN), Multilayer Perceptron (MLP), Naive Bayes (NB), and Support Vector Machine (LinSVM), AdaBoost (AB) and Random Forest (RF). We deployed the ML-based classifiers with their default scikit-learn (version 1.1.2) configurations. We created the proposed model using PyTorch. Flow features are discretized for all classifiers, as most ML algorithms perform better with discrete than continuous features. We use the discretization thresholds described in EFC's paper~\cite{efc}. Table \ref{tab:bin} details the value of the discretization thresholds.

\begin{table}
  \caption{Upper value of bins for feature discretization}
  \label{tab:bin}
  \resizebox{\linewidth}{!}{
  \begin{tabular}{ll}
    \toprule
    Feature&Upper limit of each bin\\
    \hline
    Duration& 0.001, 0.002, 0.003, 0.004, 0.005, 0.006, 0.01, 0.04, 1, 10, 100, $\infty$\\
    Protocol&TCP, UDP, GRE, ICMP, IGMP\\
    Src Pt&50, 60, 100, 400, 500, 40000, 60000, $\infty$\\
    Dst Pt&50, 60, 100, 400, 500, 40000, 60000, $\infty$\\
    Bytes&50, 60, 70, 90, 100, 110, 200, 300, 400, 500, 700, 1000, 5000, $\infty$\\
    Packets&2, 3, 4, 5, 6, 7, 10, 20, $\infty$\\
    Flags&
    $\left\{\left(f_0, f_1 , f_2 , f_3 , f_4 , f_5\right) | f_i \in \{0, 1\}\right\}$\\
    \bottomrule
  \end{tabular}
  }
\end{table}

%   The reason for choosing CIDDDS-001 and CIDDS-002 datasets should be
%     explained in Section V.

We chose the CIDDS-001 and CIDDS-002 data sets to evaluate the domain adaptation of our proposal due to their recency and the inclusion of multiple network environments. Three different environments (domains): CIDDS-001 OpenStack, CIDDS-001 External Server, and CIDDS-002 OpenStack are used to evaluate the domain adaptation capability of our proposal compared to other classifiers. The classifier will be tested on anomaly detection, which is a binary classification task. Therefore, the labels will be grouped into benign and malicious. Labels within each environment are shown in Table~\ref{tab:lab}. Flows labeled {\it normal} and {\it unknown} are considered benign, while those labeled otherwise are considered malicious. The composition of the data sets used in the experiments is shown in Table~\ref{tab:dat}.

We conducted two experiments to evaluate the domain adaptation capability of our proposal against other classifiers. In the first experiment, training was performed on the entirety of flow data from the CIDDS-001 OpenStack environment with {\it CIDDS-001 internal} data set, while testing was performed on the CIDDS-001 External Server environment with {\it CIDDS-001 external} data set and CIDDS-002 OpenStack environment with the {\it CIDDS-002} data set. As label imbalance in the training data set can impair the performance of some classifiers~\cite{iot}, a second experiment in which the training data set is balanced is conducted. In the second experiment, testing was performed on the same data sets as in the first experiment. While training was performed on a subset of flow data from the CIDDS-001 OpenStack environment. This subset, called {\it CIDDS-001 balanced}, is created by keeping all the malicious flow from the CIDDS-001 OpenStack environment while reducing the number of benign flows to match its number. The benign flows were sampled in a way that preserves the order of flows from the original data set.

\subsection{Metrics}
The performance of the proposed method is measured using Accuracy, F1-score, Recall, and Precision, which are widely utilized metrics in NIDS studies~\cite{survey}. These metrics are computed from TP (True Positive, i.e., malicious traffic classified as malicious), TN (True Negative, i.e., benign traffic classified as benign), FP (False Positive, i.e., benign traffic classified as malicious), and FN (False Negative, i.e., malicious traffic classified as benign) values. The first metric, accuracy, is defined as the ratio of correctly classified flows to the total number of flows.
\begin{equation}
  \mathrm{Accuracy} =
  \frac{\mathrm{TP}+\mathrm{TN}}{\mathrm{TP}+\mathrm{TN}+\mathrm{FP}+\mathrm{FN}}
\end{equation}
Precision is defined as the ratio of correctly classified malicious flows to all of the flows classified as malicious.
  \begin{equation}
    \mathrm{Precision}=\frac{\mathrm{TP}}{\mathrm{TP}+\mathrm{FP}}
  \end{equation}
Recall is defined as the ratio of correctly classified malicious flows to all of the malicious flows. It is also referred to as Detection Rate.
  \begin{equation}
    \mathrm{Recall}=\frac{\mathrm{TP}}{\mathrm{TP}+\mathrm{FN}}
  \end{equation}
The second metric, F1-score, is the harmonic mean of precision and recall. In other words, it is a statistical technique for examining the accuracy of a system by considering both recall and precision of the system.
\begin{equation}
  \mathrm{F1\text{-}score} =
  2\times
  \frac{\mathrm{Precision}\times\mathrm{Recall}}{\mathrm{Precision}+\mathrm{Recall}}
\end{equation}

\subsection{Evaluation Results}
We implemented a BERT model~\cite{bert} with one layer, one attention head, and a hidden size of 768. During training, we use a batch size of 512 and a sequence length of 128 flows (equivalent to 65536 flows per iteration). Adam optimizer with a constant learning rate of $10^{-5}$. In both experiments, we train the BERT model with the MLM task for 400 iterations and then fine-tune freeze the parameters of the BERT model and only train the MLP classifier for 1100 iterations, finally, we unfreeze the BERT model and trained both BERT and MLP for 400 iterations. In total, the model was trained for 2000 iterations. During testing, the sequence length is increased to 1024 flows.

\begin{table}
  \caption{Evaluation results - Training performed on CIDDS-001 internal, testing performed on CIDDS-001 external}
  \label{tab:resall-external}
  \resizebox{\linewidth}{!}{
  \begin{tabular}{lcccc}
    \toprule
    &\multicolumn{4}{c}{Train CIDDS-001 internal}\\
    &\multicolumn{4}{c}{Test CIDDS-001 external}\\
    \cline{2-5}
Classifier&Accuracy&F1-score&Recall&Precision\\
    \hline
{\bf Proposal}&{\bf 0.9078}&{\bf 0.9311}&0.9120&0.9511\\
EFC&0.8659&0.9044&{\bf 0.9278}&0.8822\\
DT&0.8491&0.8800&0.8088&0.9649\\
KNN&0.7978&0.8270&0.7067&{\bf 0.9967}\\
LinSVM&0.6784&0.6940&0.5333&0.9933\\
MLP&0.4380&0.3254&0.1982&0.9087\\
NB&0.3161&0.0000&0.0000&0.9937\\
AB&0.4606&0.3812&0.2430&0.8845\\
RF&0.8177&0.8510&0.7613&0.9647\\
    \bottomrule
  \end{tabular}
  }
\end{table}
\begin{table}
  \caption{Evaluation results - Training performed on CIDDS-001 internal, testing performed on CIDDS-002}
  \label{tab:resall-2}
  \resizebox{\linewidth}{!}{
  \begin{tabular}{lcccc}
    \toprule
    &\multicolumn{4}{c}{Train CIDDS-001 internal}\\
    &\multicolumn{4}{c}{Test CIDDS-002}\\
    \cline{2-5}
Classifier&Accuracy&F1-score&Recall&Precision\\
    \hline
{\bf Proposal}
&{\bf 0.9913}&{\bf 0.8578}&{\bf 0.7531}&{\bf 0.9962}\\
EFC
&0.9084&0.3317&0.6534&0.2223\\
DT
&0.9880&0.7948&0.6655&0.9864\\
KNN
&0.9879&0.7924&0.6656&0.9789\\
LinSVM
&0.9503&0.1042&0.0830&0.1400\\
MLP
&0.9867&0.7722&0.6479&0.9554\\
NB
&0.9638&0.0006&0.0003&0.0072\\
AB
&0.9837&0.7148&0.5873&0.9131\\
RF
&0.9881&0.7961&0.6670&0.9871\\
    \bottomrule
  \end{tabular}
  }
\end{table}

Table~\ref{tab:resall-external} and Table~\ref{tab:resall-2} show the results of the first experiment, where training was performed on CIDDS-001 internal. The proposed method presents the best performance in terms of accuracy (0.9078 and 0.9913) and F1-score (0.9311 and 0.8578) in both CIDDS-001 external and CIDDS-002 test sets. EFC ranks second in the CIDDS-001 external test set both in terms of accuracy (0.8659) and F1-score (0.9044). However, EFC performs the worst in terms of accuracy (0.9084) and the third worst in terms of the F1-score (0.3317) in the CIDDS-002 test set.

\begin{table}
  \caption{Evaluation results - Training performed on CIDDS-001 balanced, testing performed on CIDDS-001 external}
  \label{tab:resbal-external}
  \resizebox{\linewidth}{!}{
  \begin{tabular}{lcccc}
    \toprule
    &\multicolumn{4}{c}{Train CIDDS-001 balanced}\\
    &\multicolumn{4}{c}{Test CIDDS-001 external}\\
    \cline{2-5}
Classifier&Accuracy&F1-score&Recall&Precision\\
    \hline
{\bf Proposal}&{\bf 0.8581}&{\bf 0.9002}&{\bf 0.9358}&0.8673\\
EFC&0.8566&0.8985&0.9278&0.8710\\
DT&0.8496&0.8805&0.8098&0.9646\\
KNN&0.8128&0.8434&0.7374&{\bf 0.9850}\\
LinSVM&0.8123&0.8631&0.8650&0.8612\\
MLP&0.6671&0.6910&0.5441&0.9463\\
NB&0.2279&0.0282&0.0164&0.1013\\
AB&0.4742&0.4268&0.2862&0.8386\\
RF&0.8357&0.8679&0.7893&0.9639\\
    \bottomrule
    \hline
  \end{tabular}
  }
\end{table}
\begin{table}
  \caption{Evaluation results - Training performed on CIDDS-001 balanced, testing performed on CIDDS-002}
  \label{tab:resbal-2}
  \resizebox{\linewidth}{!}{
  \begin{tabular}{lcccc}
    \toprule
    &\multicolumn{4}{c}{Train CIDDS-001 balanced}\\
    &\multicolumn{4}{c}{Test CIDDS-002}\\
    \cline{2-5}
Classifier&Accuracy&F1-score&Recall&Precision\\
    \hline
{\bf Proposal}
&0.9870&0.7717&0.6315&{\bf 0.9921}\\
EFC
&0.9088&0.3327&0.6534&0.2232\\
DT
&0.9874&0.7871&0.6676&0.9586\\
KNN
&0.9874&0.7867&0.6698&0.9532\\
LinSVM
&0.5388&0.0260&0.1767&0.0140\\
MLP
&0.9865&0.7832&{\bf 0.7021}&0.8854\\
NB
&0.7908&0.0007&0.0021&0.0004\\
AB
&0.9817&0.7170&0.6653&0.7774\\
RF
&{\bf 0.9875}&{\bf 0.7891}&0.6692&0.9613\\
    \bottomrule
    \hline
  \end{tabular}
  }
\end{table}

Table~\ref{tab:resbal-external} and Table~\ref{tab:resbal-2} show the results of the second experiment, where training was performed on CIDDS-001 balanced. The proposed method presents the best performance in terms of accuracy (0.8581) and F1-score (0.9002) for CIDDS-001 external. For CIDDS-002, RF performs the best in terms of accuracy (0.9875) and F1-score (0.7891). EFC, again, exhibits strong performance in CIDDS-001 external but underperformed in CIDDS-002. DT, KNN, LinSVM, and MLP accuracy and F1-score on CIDDS-001 external improved slightly when trained on a balanced data set compared to an imbalanced one with more benign flows, as CIDDS-001 external is slightly biased toward malicious flow. The same effect is not observed on the CIDDS-002 test set due to it being biased toward benign flow, which benefits the classifier with a low false positive rate.

The extra benign flows present in the CIDDS-001 internal data set compared to the CIDDS-001 balanced data set do not have much effect on the performance of EFC, DT, KNN, AB, and RF. However, they improved the performance of the proposed method in terms of accuracy and F1 score. The proposed method benefited the most when trained on the CIDDS-001 internal data set. We hypothesize that the extra benign flows provide more contextual information to the BERT model, improving its ability to identify benign flows, while not skewing the classification result toward the class with more samples. 

%   In Tables VI and VIII, why is the proposal proposed in this manuscript so
%     much worse than the best (0.9967 and 0.9850, respectively) in terms of
%     Precision? Please analyze the possible reasons.

In cases where the proposed model is trained with CIDDS-001 internal (Table \ref{tab:resall-external}, \ref{tab:resall-2}), its Precision is much worse than KNN, the best performing method. Our proposal detected a greater amount of anomalous flows, and as a result, increases the number of normal flows flagged as malicious (false positives), which negatively impacts Precision. Whereas KNN achieved better Precision by sacrificing the number of flows flagged as malicious, and therefore, a lot of malicious flows remained undetected (false negatives).

Overall, the proposed method, together with DT, RF, and KNN was able to maintain performance across two different test sets, exhibiting good domain adaptation capability. The results for DT, RF, and KNN reflect the good performance reported in prior researches~\cite{rf, knn}, The proposed method was able to outperform all other classifiers at both test data sets when trained on CIDDS-001 internal, and CIDDS-001 external test sets when trained on CIDDS-001 balanced.

From the test result, we notice that while EFC performs well in CIDDS-001 external (same data set used in EFC's original paper~\cite{efc}), its F1-score for the CIDDS-002 test set (not used in EFC's original paper) was significantly worse. We ran an experiment with smaller balanced test sets (with 10000 benign and 10000 malicious flows each) sampled from CIDDS-002~\cite{own}. We observed that in the smaller test setting, EFC achieved results comparable to DT, RF, and KNN. We found that EFC performed well in small and balanced test sets (as reported by Camila {\it et al.}) but its performance degrades significantly in a large imbalanced test set heavily skewered toward benign flows (the CIDDS-002 test set).

\section{Conclusion}
We proposed an NIDS that utilizes a sequence of flows as input and the BERT model. We theorize that because conventional ML-based NIDSs only use information from a single flow, they exhibit poor domain adaptation capability. The training of our proposed model consists of two steps. In the first step, we trained the BERT model to predict a flow given its context using unlabeled data. In the second step, an MLP classifier is added to the output of the BERT model and trained on labeled data. Early experimental results on CIDDS-001 and CIDDS-002 showed the proposed method achieving consistent results across different domains.

%Currently, our model extracts feature from a sequence of flows as it was captured from an observation point. These sequences, therefore, represent the overall behavior of the network, with flow from different hosts included in the same sequence. In this case, the model detects flows that deviate from the norm of the network. We plan to separate the mixed sequence into sequences of flows originating from the same host. The idea behind this is to train the model to better represent the behaviors of the hosts themselves. With this approach, the model identifies host activities patterns, and because flows from a benign host are more likely to be benign, theoretically, it can lower false alarm rate and increase detection rate.

In this study, we used a feedforward neural network to classify benign and malicious flows, which requires labeled flow to train the model. In the future, we plan to model the distribution of benign flow sequences and use the deviations from this distribution to detect anomalous flows.

%In this study, we used an MLP classifier to classify benign and malicious flows, which requires labeled flow to train the model. In the future, we plan to experiment with a statistical approach toward flow classification, with the system modeling the distribution of benign flow sequences, and deviations from this distribution indicate a higher probability of anomalous flows.

\bibliographystyle{IEEEtran}
\bibliography{ref}
\end{document}